\documentclass[aps,preprint,tightenlines,byrevtex,
               superscriptaddress,showpacs]{revtex4}
\usepackage{graphicx}
\usepackage{bm}

\begin{document}

\preprint{THEF-NYM 02.04}
\preprint{KVI-1585}

\title{Determination of the chiral coupling constants \bm{$c_3$} and
       \bm{$c_4$} in new \bm{$pp$} and \bm{$np$} partial-wave analyses}

\author{M.C.M. Rentmeester}
\affiliation{Institute for Theoretical Physics,
             University of Nijmegen,
             P.O. Box 9010,
             6500 GL Nijmegen, The Netherlands}
\author{R.G.E. Timmermans}
\affiliation{Theory Group,
             KVI, University of Groningen,
             Zernikelaan 25,
             9747 AA Groningen, The Netherlands}
\author{J.J. de Swart}
\affiliation{Institute for Theoretical Physics,
             University of Nijmegen,
             P.O. Box 9010,
             6500 GL Nijmegen, The Netherlands}

\date{\today}

\begin{abstract}
As a first result of two new partial-wave analyses, one of the $pp$
and another one of the $np$ scattering data below 500 MeV, 
we report a study of the long-range chiral two-pion exchange interaction 
which contains the chiral coupling constants $c_1$, $c_3$, and $c_4$. 
By using as input a theoretical value for $c_1$ we are able to determine
in $pp$ as well as in $np$ scattering accurate values for $c_3$ and $c_4$.
The values determined from the $pp$ data and independently from the $np$
data are in very good agreement, indicating the correctness of the chiral
two-pion exchange interaction. The weighted averages are
                   $c_3=-4.78(10)$/GeV and
                   $c_4= 3.96(22)$/GeV,
where the errors are statistical. The value of $c_3$ is best determined
from the $pp$ data and that of $c_4$ from the $np$ data.
\end{abstract}
\maketitle

\section*{Introduction}
It is beyond doubt that the longest-range strong two-nucleon ($N\!N$)
interaction is the one-pion exchange (OPE) force. Despite more than
50 years of research, the nature of the medium-range $N\!N$ interaction
is not so well understood. What seems clear is that it contains:
($i$) A strongly attractive central force,
($ii$) an isospin-dependent tensor force opposite in sign to OPE,
and ($iii$) a rather strongly attractive spin-orbit force. 
It was discovered in the early sixties that all these features follow 
naturally from the exchange of scalar and vector mesons, which led to
the development of the one-boson exchange (OBE) model of the $N\!N$
interaction. The role of the two-pion exchange (TPE) interaction and
its interplay with the exchange of heavy mesons that decay into two
pions has for a long time remained elusive.

In recent years, however, the situation has improved. The derivation of
at least the long- and medium-range nuclear forces can be formulated in a
model-independent manner by a systematic expansion of the chiral Lagrangian
of QCD~\cite{Wei90,Ord92,Fri94,Kai97,Kap98,Epe98,Ren99}. In particular, the
long-range TPE interaction can be derived unambiguously, where the effects
of the exchange of broad heavy mesons are incorporated in effective
low-energy chiral coupling constants. Most importantly, chiral symmetry
and its breaking are correctly implemented in this approach.

In Ref.~\cite{Ren99} we studied this long-range chiral two-pion exchange
($\chi$TPE) interaction in an energy-dependent partial-wave analysis (PWA)
of the proton-proton ($pp$) scattering data below 350 MeV. 
The presence of $\chi$TPE in the long-range $pp$ force was demonstrated,
and the chiral coupling constants $c_3$ and $c_4$ were determined 
from the $pp$ data. In this paper we address the question whether
the same $\chi$TPE force allows also a good description of the
neutron-proton ($np$) scattering data below 500 MeV. Moreover, we
present new, precise determinations of the chiral coupling constants
$c_3$ and $c_4$ from the $pp$ and $np$ data separately. Accurate values
of these chiral coupling constants are an important input in calculations
of, for instance, the two-pion exchange three-nucleon force~\cite{Fri98}.

\section*{Partial-wave analysis}
Because of the high quality of the $pp$ data base all the $pp$ phase
shifts with orbital angular momentum $\ell\leq 4$ can be determined 
accurately in an energy-dependent PWA of the $pp$ data below 350 MeV.
An analysis of the $np$ data, however, is much more difficult, 
because not only the $I=1$ phase shifts but also the $I=0$ phase 
shifts contribute. 
Moreover, the $np$ data base, while extensive, is by far not as accurate 
and varied as the $pp$ data base~\cite{NNOnL}. 
In a PWA of only the $np$ data below 350 MeV it has always 
been impossible to determine all the important $np$ phase shifts.
Therefore, the standard practice has been to take the $I=1$ phase
shifts, with the exception of the $^1S_0$ phase shift, from the $pp$ PWA,
with or without corrections for the Coulomb interaction and/or the
$\pi^+$-$\pi^0$ mass difference in OPE.
Since it has long been known that there is a sizable charge-independence 
breaking (CIB) in the $^1S_0$ phase shifts, the $^1S_0$ $np$ phase shift 
is always fitted independently of the $^1S_0$ $pp$ phase shift.

This approach to $np$ PWA was also followed in the past by the Nijmegen
group. In 1993 the results of the first Nijmegen $pp$ and $np$ PWA's below
350 MeV were published in Ref.~\cite{Sto93}. 
An attempt at that time to extract all the important $np$ phase shifts, 
both $I=0$ and $I=1$, from the $np$ data base below 350 MeV failed, 
although it was possible to determine the $^3P$ $np$ phase shifts
when the $I=1$ waves for $\ell>1$ were taken over from the
$pp$ PWA93~\cite{Klo93}.

It has been customary to perform $N\!N$ PWA's without inelasticities up
to 350 MeV, although pion production starts already at 280 MeV. 
It can be shown that the inclusion of inelasticities in the $pp$ PWA 
below 350 MeV improves the $\chi^2_{min}$ slightly. 
Already some time ago~\cite{Kok93} the Nijmegen $pp$ PWA was 
extended to energies far above the pion-production thresholds, with the 
inclusion of inelasticities. 
When, in 1994, the $np$ PWA was extended to 500 MeV, it turned out to be 
possible, for the first time, to determine uniquely all the important $np$
phase shifts, both $I=0$ and $I=1$, from the $np$ data alone~\cite{Klo94}.
Such a separate PWA of the $np$ data, without input from the $pp$ PWA for 
the $I=1$ waves, is in principle more model independent. 
A comparison between the phase shifts from such an independent $np$ PWA 
and the corresponding phase shifts from the $pp$ PWA 
provides information about possible CIB in the $I=1$ waves.

The Nijmegen energy-dependent PWA's can be used as a tool to study the
long-range $N\!N$ interaction~\cite{Ren99}. 
The long-range forces are included exactly, in order to ensure that the 
partial-wave amplitudes acquire the proper fast energy dependence 
from the nearby left-hand singularies due to these long-range forces, 
while the short-range interactions (more remote left-hand singularities), 
responsible for a much slower energy dependence, are parametrized. 
This strategy is implemented by solving the Schr\"odinger equation with 
an energy-dependent boundary condition (BC) at some $r=b$ and for $r>b$ 
the long-range $N\!N$ interaction. 
This long-range force contains the electromagnetic interaction 
({\it i.e.}, in the $pp$ case the improved Coulomb~\cite{Aus83},
the magnetic-moment~\cite{Sto90}, and the vacuum-polarization~\cite{Dur57}
interactions, 
and in the $np$ case the magnetic-moment interaction~\cite{Sto90}), 
the OPE interaction, in the $np$ case also the pion-photon ($\pi$-$\gamma$) 
exchange interaction~\cite{Kol98}, and the long-range part of the $\chi$TPE
interaction~\cite{Ren99}.
The BC is parametrized as an analytic function of energy, and the parameters,
representing ``short-range physics,'' are determined from a fit to the data.
The option also exists to fit simultaneously some of the parameters in the 
long-range interactions, {\it viz}. the pion-nucleon coupling
constants~\cite{Ber87,Tim93,Swa97} and/or
the chiral coupling constants $c_i$ ($i=1,3,4$) in $\chi$TPE~\cite{Ren99}.

The new $pp$ and $np$ PWA's that we discuss here will be referred
to as $\chi$PWA03. They differ from the old PWA93 in several aspects:
($i$)   The energy range is extended from 350 to 500 MeV.
        Instead of the 1787 $pp$ data and 2514 $np$ data in PWA93 we
        now have 5109 $pp$ data and 4786 $np$ data~\cite{Dat01}.
($ii$)  All the $np$ phase shifts can be determined from the $np$ data
        alone, instead of taking the $I=1$ phases from the $pp$ PWA and
        correcting them.
($iii$) Inelasticities are taken into account.
($iv$)  For $r>b$ a different non-OPE strong interaction is taken.
        In PWA93 the heavy-meson exchanges of the Nijmegen soft-core
        OBE potential~\cite{Nag78} were used. 
        Motivated by the success of Ref.~\cite{Ren99}, where an excellent
        description of the high-quality $pp$ data base below 350 MeV was
        obtained, we use in $\chi$PWA03 the $\chi$TPE potential.
($v$)   A minor difference between PWA93 and $\chi$PWA03 is that we take
        here $b=1.6$ fm, while in PWA93 $b=1.4$ fm was used; 
        in Ref.~\cite{Ren99} we used both $b=1.4$ fm and $b=1.8$ fm. 

Details of $\chi$PWA03 (data, phase shifts, {\it etc.}) will be presented 
elsewhere~\cite{Ren03}, here we focus on testing the long-range $\chi$TPE 
interaction in the $pp$ and $np$ systems below 500 MeV.

\section*{Chiral two-pion exchange potential}
The $\chi$TPE potential can be derived by a systematic expansion of the
effective chiral Lagrangian~\cite{Wei90,Ord92,Fri94,Kai97,Kap98,Epe98,Ren99}.
The form that is appropriate for use in the relativistic Schr\"odinger
equation and that is consistent with our choice of 
including the minimal-relativity factor $M/E$ in the OPE potential
is specified in Ref.~\cite{Ren99} (see also Ref.~\cite{Fri94}).

The leading-order $\chi$TPE potential consists of the static planar- and
crossed-box TPE diagrams, calculated with the derivative (pseudovector)
$N\!N\pi$ Lagrangian, and the triangle and football diagrams with the
nonlinear $N\!N\pi\pi$ Weinberg-Tomozawa (WT) seagull vertices. It contains
isospin-independent spin-spin and tensor terms and an isospin-dependent
central term.

In subleading order, next to nonstatic corrections to the planar and
box diagrams, additional triangle diagrams appear which contain three
new $N\!N\pi\pi$ interactions~\cite{Ord92}.
The corresponding chiral coupling constants are denoted by $c_i$ ($i=1,3,4$).
(Unfortunately, they are not scaled to obtain dimensionless numbers and
their values are conventionally given in GeV$^{-1}$.)
They are defined by the following terms in the chiral Lagrangian density:
\begin{eqnarray}
  {\mathcal L} & = & -\overline{N}\left[\:
                     8 c_1 D^{-1}m_\pi^2\vec{\pi}^2/F_\pi^2 +
                     4 c_3\,\vec{D}_\mu\!\cdot\!\vec{D}^\mu \right.
  \nonumber \\ &   & \left. + \,
                     2 c_4\,\sigma_{\mu\nu}\,\vec{\tau}\cdot
                     \vec{D}^\mu\!\times\!\vec{D}^\nu \:\right]N
  \ , \label{eq:c134}
\end{eqnarray}
where $F_\pi\simeq 185$ MeV is the pion decay constant,
$D=1+\vec{\pi}^2/F_\pi^2$, and the chiral-covariant derivative of
the pion field $\vec{\pi}$ is
$\vec{D}^\mu = D^{-1}\partial^\mu\vec{\pi}/F_\pi$.
The $c_3$- and $c_4$-terms are manifestly chiral invariant. The $c_1$-term,
which is proportional to $m_\pi^2$, violates chiral symmetry explicitly and
is related to the much-discussed pion-nucleon sigma term~\cite{Gas91}.
Using the rationalized pseudovector pion-nucleon coupling constant $f^2$
the relation reads~\cite{Ber92}
\begin{equation}
    c_1 = -\left[ \frac{\sigma}{4m_\pi^2} +
           \frac{9}{16}\frac{f^2}{m_s^2}\,m_\pi \right] \ , \label{eq:sigma}
\end{equation}
where $m_\pi=138.04$ MeV is the average pion mass, and $m_s\equiv m_{\pi^+}$
is the scaling mass conventionally introduced to make $f$ dimensionless.
Eq.~(\ref{eq:sigma}) holds in order ${\cal O}(q^3)$ in the chiral expansion
in small momenta $q$ and the pion mass~\cite{Ber92}.
An additional $c_2$-term is not given in Eq.~(\ref{eq:c134}), 
since it does not contribute to the $\chi$TPE potential to subleading order. 
However, it does contribute to the isoscalar $\pi N$ scattering amplitude
at the same order as $c_1$ and $c_3$. 

In subleading order the $\chi$TPE potential gets contributions to the
central, spin-spin, tensor, and spin-orbit potentials ({\it cf.} Table 1 
in Ref.~\cite{Ren99}). 
Important components are: 
($i$) A strong isospin-independent central attraction due to the $c_3$-term,
($ii$) an isospin-dependent tensor force opposite in sign to OPE due to the
$c_4$-term, and
($iii$) an attractive isospin-independent spin-orbit force from nonstatic 
terms of the planar- and crossed-box diagrams.
The values of the $c_i$'s are not fixed by chiral symmetry and must 
be determined from the experimental $\pi N$ or $N\!N$ scattering data.

The long-range $\chi$TPE potential derived in the framework of the 
effective chiral Lagrangian is completely model independent. 
Any dynamical {\em model}~\cite{Rij91,Rob01} for the TPE $N\!N$ interaction, 
containing {\em e.g.} the $\varepsilon$ (or ``$\sigma$'') and $\varrho$ 
mesons, the pomeron, and/or $N$- and $\Delta$-isobars, has to reduce to
this form for large $r$. These models should also predict values for the
$c_i$'s consistent with the determinations from the $\pi N$ and $N\!N$
scattering data.

The breaking of charge-independence due to the $\pi^+$-$\pi^0$ mass difference
in the OPE potential is taken into account exactly, as it was already
in PWA93~\cite{Sto93}. In the $\chi$TPE potential we include the terms
linear in the $\pi^+$-$\pi^0$ mass difference, following Ref.~\cite{Jim99}.
One charge-independent pion-nucleon coupling constant~\cite{Swa97} is
used in both the OPE and the $\chi$TPE potentials. In the long-range
interaction for $r>b$ only the chiral coupling constants $c_i$ ($i=1,3,4$) 
remain then to be determined.

\section*{Results}
In our previous study~\cite{Ren99} of $\chi$TPE it turned out that $c_1$
could not be determined accurately from the $pp$ data base below 350 MeV.
When we fitted $c_1$, $c_3$, and $c_4$ simultaneously, we found
$c_1=-4.4(3.4)$/GeV, where the error is statistical. A strong correlation
was obtained between the values of $c_1$ and $c_3$. Therefore, we used
Eq.~(\ref{eq:sigma}) to fix the value of $c_1$. Assuming that the sigma
term has the low value $\sigma=35(5)$ MeV~\cite{Gas91,Tim97}, 
Eq.~(\ref{eq:sigma}) gives $c_1=-0.76(7)$/GeV, where the error is
theoretical. We used the central value $c_1=-0.76$/GeV as input 
in the PWA, and determined in Ref.~\cite{Ren99} the values
$c_3=-5.08(28)$/GeV and $c_4=4.70(70)$/GeV, where the errors are
statistical. One notes that from the $pp$ data below 350 MeV the
value of $c_3$ could be extracted rather precisely, while $c_4$
was pinned down less accurately.

The value extracted for $c_1$ from the data below 500 MeV would also not
be accurate enough to shed light on the value of the sigma term, since the
statistical error for $c_1$ obtained in Ref.~\cite{Ren99} would have to be
reduced at least by a factor of about 20.
We therefore decided to use also here the value $c_1=-0.76$/GeV as input 
value, and to determine $c_3$ and $c_4$ from direct fits to the $pp$ data
and independently also from fits to the $np$ data.

We analyzed 5109 $pp$ data below 500 MeV using 33 BC parameters, and we
reached $\chi^2_{min}=5184.3$. The optimal values for $c_3$ and $c_4$ and
their (1 s.d.) errors as determined from these $pp$ data are:
\begin{eqnarray}
  c_3 & = & \left[-4.78(11)+80(f^2-0.0755)\right]/{\rm GeV} \ , \nonumber \\
  c_4 & = & \left[3.92(52)+260(f^2-0.0755)\right]/{\rm GeV} \ ,
\end{eqnarray}
where also the dependence on the $N\!N\pi$ coupling constant $f^2$
is displayed. The correlation parameter is $\varrho=-0.47$.
These values for $c_3$ and $c_4$ are consistent with and more accurate
than those found in Ref.~\cite{Ren99} from the $pp$ data below 350 MeV. 
The errors are statistically only.
Systematic errors are difficult to assess and require further study.

\begin{figure}
\begin{center}
\includegraphics{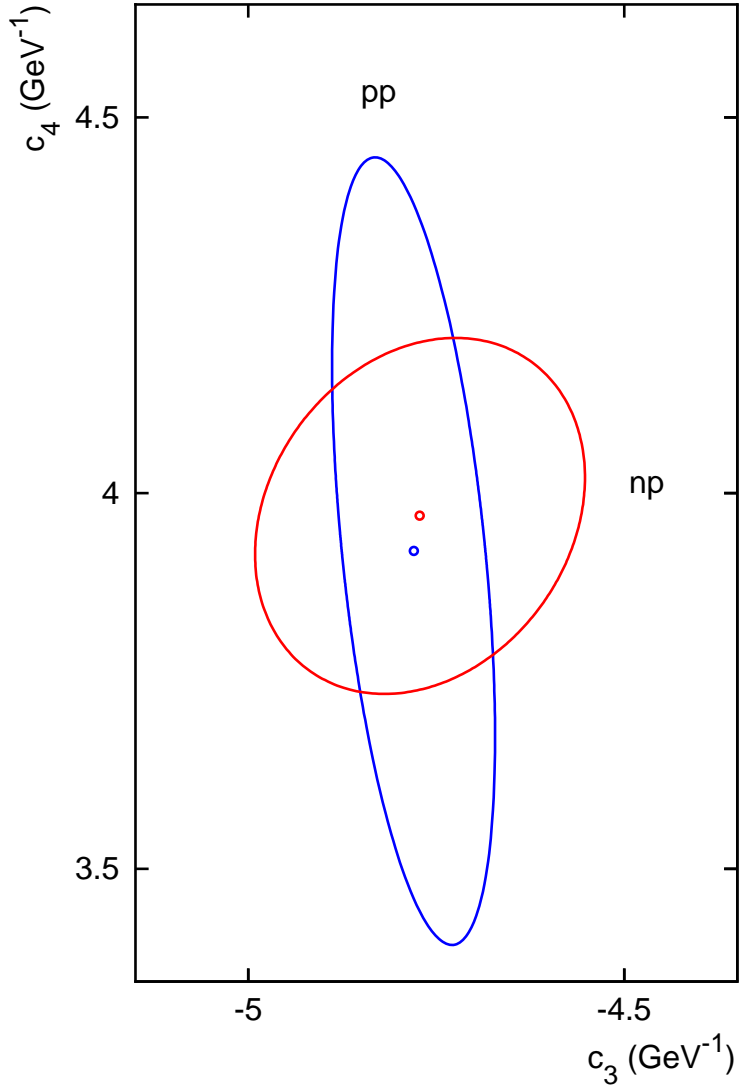}
\end{center}
\caption{\label{fig:1} Ellipses of constant $\chi^2$ in the $(c_3,c_4)$
         plane. Shown are the $\chi^2=\chi^2_{min}+1$ ellipses in the
         $pp$ PWA and in the $np$ PWA. The centers of the ellipses
         correspond to the minima in $\chi^2$.}
\end{figure}

For $np$ scattering we analyzed 4786 data below 500 MeV.
In this case we needed 40 BC parameters and reached $\chi^2_{min}=4806.2$.
The chiral coupling constants, their statistical errors, and their 
dependence on the $N\!N\pi$ coupling constant are in this $np$ case: 
\begin{eqnarray}
  c_3 & = & \left[-4.77(22)+100(f^2-0.0755)\right]/{\rm GeV} \ , \nonumber \\
  c_4 & = & \left[ 3.97(24)+ 40(f^2-0.0755)\right]/{\rm GeV} \ .
\end{eqnarray}
The correlation parameter is $\varrho=0.22$.
In Fig.~\ref{fig:1} we show the results for $c_3$ and $c_4$, for
$f^2=0.0755$. Plotted are the positions of the $\chi^2$-minima and 
the $\chi^2=\chi^2_{min}+1$ ellipses in the $(c_3,c_4)$ plane, 
both for the $pp$ and the $np$ case. (These ellipses, of course, 
are determined with optimalization of all the BC parameters.)

The values of $c_3$ and $c_4$ determined from the $pp$ and from the $np$
data are in good agreement. The value for $c_3$ determined from the $pp$
data is more than twice as accurate as the value from the $np$ data,
while for $c_4$ the situation is reversed: the value from the $np$ data
is twice as accurate as the value from the $pp$ data.

We also determined the weighted averages with errors. We get
\begin{eqnarray}
  c_3 & = & \left[-4.78(10)+84(f^2-0.0755)\right]/{\rm GeV} \ , \nonumber \\
  c_4 & = & \left[ 3.96(22)+79(f^2-0.0755)\right]/{\rm GeV} \ .
\end{eqnarray}
These are our best values, following from all the $pp$ and $np$ scattering
data below 500 MeV, which amounts to a total of almost ten thousand $N\!N$
data. In Table~\ref{tab:1} we list our results for $c_3$ and $c_4$, for
$f^2=0.0755$.

\section*{Discussion and summary}
We have determined accurate values for the chiral coupling constants
$c_3$ and $c_4$ from the $pp$ and the $np$ scattering data below 500 MeV.
The values for the $c_i$'s ($i=1,2,3,4$) can also be determined from
PWA's of the $\pi N$ scattering data 
by fitting the amplitudes predicted by (heavy-baryon) chiral perturbation
theory ($\chi$PT) to the $\pi N$ scattering amplitudes obtained 
from these PWA's. In the several such determinations 
(for a discussion of the status see Ref.~\cite{Epe02}) the
value of $c_3$ is found to lie in the range between $-4.70$ and $-6.19$,
and the value of $c_4$ in the range between $3.25$ and $4.12$. 
The values that are found for $c_3$ and $c_4$ depend on the order
in $\chi$PT of the amplitudes, as well as on the specific $\pi N$ PWA
that is used. They also depend on what value is used for the sigma term,
because this value fixes the value of $c_1$. In $\chi$PT the isovector
$\pi N$ amplitudes can be predicted more accurately than the isoscalar
amplitudes, because in leading order the latter are zero. Therefore,
$c_4$ can probably be pinned down better than $c_3$. The value of $c_3$
is moreover strongly correlated with the values of $c_1$ and $c_2$.
In Table~\ref{tab:1} we also listed the values obtained for the $c_i$'s in
two $\pi N$ analyses (to order ${\cal O}(q^3)$ in $\chi$PT) that assume,
like us, an acceptably low value for the sigma term.
(From Ref.~\cite{Ber97} we list only the $c_i$'s corresponding 
to one of the fits with $f^2=0.076$.)

\begin{table}
\caption{\label{tab:1}
         Comparison of the chiral coupling constants $c_i$ ($i=1,3,4$)
         (in units of 1/GeV) from different analyses. The $\pi N$ results
         correspond to analyses to order ${\cal O}(q^3)$ in $\chi$PT.
         The input values of the sigma term are in MeV.
         (For a discussion of the meaning of the errors, see the text.)}
\begin{ruledtabular}
\begin{tabular}{lc|cccc}
        &   Ref.        & $\sigma$ & $c_1$ &  $c_3$ & $c_4 $  \\ \hline
$\pi N$ & \cite{Ber97}  & $45(8)$          &    $-0.91(9)  $
                        & $-5.16(25)$      &    $ 3.63(10) $  \\
$\pi N$ & \cite{But00}  & $40(8)$          &    $-0.81(12) $
                        & $-4.70(1.16)$    &    $ 3.40(04) $  \\ 
$pp$    & \cite{Ren99}  & $35(5)$          &    $-0.76(7)  $
                        & $-5.08(28)$      &    $ 4.70(70) $  \\
$pp$    & This work     & $35(5)$          &    $-0.76(7)  $
                        & $-4.78(11)$      &    $ 3.92(52) $  \\
$np$    & This work     & $35(5)$          &    $-0.76(7)  $
                        & $-4.77(22)$      &    $ 3.97(24) $  \\
$N\!N$  & This work     & $35(5)$          &    $-0.76(7)  $
                        & $-4.78(10)$      &    $ 3.96(22) $
\end{tabular}
\end{ruledtabular}
\end{table}

A problem with these determinations of the $c_i$'s from the
$\pi N$ scattering data
is that they are not determined directly from the data, but from
fitting to the amplitudes of existing PWA's that have no reliable errors.
For instance, in several analyses the amplitudes of the about 25-year-old
Karlsruhe-Helsinki dispersion analysis were used. The amplitudes of that
PWA have no associated errors and, 
what is worse, are in disagreement with the modern-day $\pi N$ data base.
That analysis produced the high value $f^2=0.079$ for the pion-nucleon
coupling constant~\cite{Swa97}.
The resulting errors on the $c_i$'s determined from the $\pi N$ data,
therefore, do not reflect the statistics of the data base, 
but are essentially rather arbitrary estimates.

Our results correspond to a long-range interaction that includes the 
leading and subleading $\chi$TPE diagrams. 
Higher-order corrections for $\chi$TPE and the leading three-pion exchange 
diagrams have been calculated by Kaiser~\cite{Kai00}, and can in principle 
be included as well. Work along these lines is continuing.

In summary, the long-range part of the $\chi$TPE potential was included
in energy-dependent PWA's of the $pp$ and the $np$ scattering data below
500 MeV. Good fits to the data were obtained. 
In the $np$ PWA all the phase-shift parameters could be determined without 
input from the $pp$ system. 
We conclude that OPE plus $\chi$TPE provides a high-quality long-range 
strong two-nucleon interaction. 
Accurate values for the chiral coupling constants $c_3$ and $c_4$ of chiral 
perturbation theory were obtained from the $pp$ and the $np$ data separately. 
The values agree very well with each other,
and they are also in good agreement with the range of values obtained from
pion-nucleon scattering amplitudes.
We consider this agreement to be experimental evidence that the $\chi$TPE
interaction, as predicted by chiral perturbation theory, is correct.

\begin{acknowledgments}
We thank Th.A. Rijken for helpful discussions.
The research of R.G.E.T. was made possible by a fellowship of 
the Royal Netherlands Academy of Arts and Sciences (KNAW).
\end{acknowledgments}


\begin{thebibliography}{XXX99}
\bibitem{Wei90} S. Weinberg, Phys. Lett. B {\bf 251}, 288 (1990);
                Nucl. Phys. {\bf B363}, 3 (1991); Phys. Lett. B
                {\bf 295}, 114 (1992).
\bibitem{Ord92} C. Ord\'o\~nez and U. van Kolck, Phys. Lett. B
                {\bf 291}, 459 (1992); C. Ord\'o\~nez, L. Ray, and
                U. van Kolck, Phys. Rev. Lett. {\bf 72}, 1982 (1994);
                Phys. Rev. C {\bf 53}, 2086 (1996).
\bibitem{Fri94} J.L. Friar and S.A. Coon,
                Phys. Rev. C {\bf 49}, 1272 (1994);
                J.L. Friar, Phys. Rev. C {\bf 60}, 034002 (1999).
\bibitem{Kai97} N. Kaiser, R. Brockmann, and W. Weise,
                Nucl. Phys. {\bf A625}, 758 (1997);
                N. Kaiser, S. Gerstend\"orfer, and W. Weise,
                Nucl. Phys. {\bf A637}, 395 (1998).
\bibitem{Kap98} D.B. Kaplan, M.J. Savage, and M.B. Wise,
                Nucl. Phys. {\bf B534}, 329 (1998);
                S. Fleming, T. Mehen, and I.W. Stewart,
                Nucl. Phys. {\bf A677}, 313 (2000).
\bibitem{Epe98} E. Epelbaum, W. Gl\"ockle, and U.-G. Meissner,
                Nucl. Phys. {\bf A637}, 107 (1998);
                Nucl. Phys. {\bf A671}, 295 (2000).
\bibitem{Ren99} M.C.M. Rentmeester, R.G.E. Timmermans, J.L. Friar, and
                J.J. de Swart, Phys. Rev. Lett. {\bf 82}, 4992 (1999).
\bibitem{Fri98} J.L. Friar, D. H\"uber, and U. van Kolck,
                Phys. Rev. C {\bf 59}, 53 (1999).
\bibitem{NNOnL} NN-OnLine facility,
                {\tt http://NN-OnLine.sci.kun.nl}.
\bibitem{Sto93} V.G.J. Stoks, R.A.M. Klomp, M.C.M. Rentmeester, and
                J.J. de Swart, Phys. Rev. C {\bf 48}, 792 (1993). This
                PWA93 has been made available on the web~\cite{NNOnL}.
\bibitem{Klo93} R.A.M.M. Klomp, unpublished (1993).
\bibitem{Kok93} J.L. de Kok, Ph.D. Thesis, University of Nijmegen,
                Nijmegen, The Netherlands (1993).
\bibitem{Klo94} R.A.M.M. Klomp, J.-L. de Kok, M.C.M. Rentmeester,
                Th.A. Rijken, and J.J. de Swart, in: {\it Few-Body
                Problems in Physics, Williamsburg 1994}, AIP
                Conference Proceedings 334, edited by F. Gross
                (AIP Press, 1995), p. 367.
\bibitem{Aus83} G.J.M. Austen and J.J. de Swart,
                Phys. Rev. Lett. {\bf 50}, 2039 (1983).
\bibitem{Sto90} V.G.J. Stoks and J.J. de Swart,
                Phys. Rev. C {\bf 42}, 1235 (1990).
\bibitem{Dur57} L. Durand III, Phys. Rev. {\bf 108}, 1597 (1957).
\bibitem{Kol98} U. van Kolck, M.C.M. Rentmeester,
                J.L. Friar, T. Goldman, and J.J. de Swart,
                Phys. Rev. Lett. {\bf 80}, 4386 (1998).
\bibitem{Ber87} J.R. Bergervoet, P.C. van Campen,
                Th.A. Rijken, and J.J. de Swart,
                Phys. Rev. Lett. {\bf 59}, 2255 (1987);
                R.A.M. Klomp, V.G.J. Stoks, and J.J. de Swart,
                Phys. Rev. C {\bf 44}, R1258 (1991);
                R. Timmermans, Th.A. Rijken, and J.J. de Swart,
                Phys. Rev. Lett. {\bf 67}, 1074 (1991).
\bibitem{Tim93} V. Stoks, R. Timmermans, and J.J. de Swart,
                Phys. Rev. C {\bf 47}, 512 (1993).
\bibitem{Swa97} J.J. de Swart,
                M.C.M. Rentmeester, and R.G.E. Timmermans,
                $\pi N$ Newsletter {\bf 13}, 96 (1997),
                {\tt nucl-th/9802084}.
\bibitem{Dat01} In $\chi$PWA03 the $N\!N$ database of January 2001
                is used~\cite{NNOnL}.
\bibitem{Nag78} M.M. Nagels, Th.A. Rijken, and J.J. de Swart,
                Phys. Rev. D {\bf 17}, 768 (1978).
\bibitem{Ren03} M.C.M. Rentmeester, R.G.E. Timmermans, and
                J.J. de Swart, in preparation.
\bibitem{Gas91} J. Gasser, H. Leutwyler, and M.E. Sainio,
                Phys. Lett. B {\bf 253}, 252, 260 (1991).
\bibitem{Ber92} V. Bernard, N. Kaiser, J. Kambor, and U.-G.
                Meissner, Nucl. Phys. {\bf B388}, 315 (1992).
\bibitem{Rij91} Th.A. Rijken,
                Ann. Phys. (N.Y.) {\bf 208}, 253 (1991);
                Th.A. Rijken and V.G.J. Stoks,
                Phys. Rev. C {\bf 54}, 2851, 2869 (1996).
\bibitem{Rob01} M.R. Robilotta, Phys. Rev. C {\bf 63}, 044004 (2001).
\bibitem{Jim99} J.L. Friar and U. van Kolck,
                Phys. Rev. C {\bf 60}, 034006 (1999).
\bibitem{Tim97} R.G.E. Timmermans,
                $\pi N$ Newsletter {\bf 13}, 80 (1997).
\bibitem{Epe02} E. Epelbaum, A. Nogga, W. Gl\"ockle, H. Kamada, U.-G.
                Meissner, and H. Witala, Eur. Phys. J. {\bf A15}, 543 (2002).
\bibitem{Ber97} V. Bernard, N. Kaiser, and U.-G. Meissner,
                Nucl. Phys. {\bf A615}, 483 (1997).
\bibitem{But00} P. B\"uttiker and U.-G. Meissner,
                Nucl. Phys. {\bf A668}, 97 (2000).
\bibitem{Kai00} N. Kaiser,
                Phys. Rev. C {\bf 61}, 014003 (2000);
                {\it ibid}. {\bf 62}, 024001 (2000);
                {\bf 63}, 044010 (2001);
                {\bf 64}, 057001 (2001);
                {\bf 65}, 017001 (2002).
\end{thebibliography}
\end{document}